\documentclass[%
 reprint,
 superscriptaddress,
 amsmath,amssymb,
 aps,
]{revtex4-2}

\usepackage{graphicx}
\usepackage{dcolumn}
\usepackage{bm}

\usepackage[version=4]{mhchem}
\usepackage{siunitx}

\makeatletter
\def\maketitle{
\@author@finish
\title@column\titleblock@produce
\suppressfloats[t]}
\makeatother

\begin{document}

\preprint{APS/123-QED}

\title{Focusing Terawatt-Scale Lasers Using Holographic Gaseous Lenses}

\author{Devdigvijay Singh}
\affiliation{Department of Mechanical Engineering, Stanford University, Stanford, 94305, CA, USA}

\author{Ke Ou}
\affiliation{Department of Mechanical Engineering, Stanford University, Stanford, 94305, CA, USA}

\author{Sida Cao}
\affiliation{Department of Mechanical Engineering, Stanford University, Stanford, 94305, CA, USA}

\author{Victor M. Perez-Ramirez}
\affiliation{Department of Mechanical Engineering, Stanford University, Stanford, 94305, CA, USA}

\author{Harsha Rajesh}
\affiliation{Department of Mechanical Engineering, Stanford University, Stanford, 94305, CA, USA}

\author{Debolina Chakraborty}
\affiliation{Department of Mechanical Engineering, Stanford University, Stanford, 94305, CA, USA}

\author{Elaine Koh}
\affiliation{Department of Mechanical Engineering, Stanford University, Stanford, 94305, CA, USA}

\author{Caleb Redshaw}
\affiliation{Department of Mechanical Engineering, Stanford University, Stanford, 94305, CA, USA}

\author{Pelin Dedeler}
\affiliation{Department of Mechanical Engineering, Stanford University, Stanford, 94305, CA, USA}

\author{Albertine Oudin}
\affiliation{Lawrence Livermore National Laboratory, Livermore, 94551, CA, USA}

\author{Michelle M. Wang}
\affiliation{Department of Electrical Engineering, Princeton University, Princeton, 08544, NJ, USA}

\author{Julia M. Mikhailova}
\affiliation{Department of Mechanical and Aerospace Engineering, Princeton University, Princeton, 08544, NJ, USA}

\author{Livia Lancia}
\affiliation{LULI, CNRS, CEA, Sorbonne Universit\'e, \'Ecole Polytechnique, Palaiseau, F-91128, France}

\author{Caterina Riconda}
\affiliation{LULI, Sorbonne Universit\'e, CNRS, \'Ecole Polytechnique, CEA, Paris, F-75252, France}

\author{Pierre Michel}
\affiliation{Lawrence Livermore National Laboratory, Livermore, 94551, CA, USA}

\author{Matthew R. Edwards}
\email{mredwards@stanford.edu}
\affiliation{Department of Mechanical Engineering, Stanford University, Stanford, 94305, CA, USA}

\date{\today}

\begin{abstract}
The capabilities of the world's highest energy and peak-power pulsed lasers are limited by optical damage, and further advances in high-intensity laser science will require optics that are substantially more robust than existing components. 
We describe here the experimental demonstration of off-axis diffractive gaseous lenses capable of withstanding extreme laser fluence and immune to cumulative damage. 
We used less than 8~mJ of energy from interfering ultraviolet laser pulses to holographically write millimeter-scale diffractive gas lenses into an ozone, oxygen, and carbon-dioxide gas mixture. These lenses allowed us to focus, defocus, and collimate 532-nm, 210-mJ nanosecond laser pulses at fluences up to 35 J/cm\textsuperscript{2} and 800-nm, 35-fs, 28-mJ femtosecond pulses at intensities up to 55~TW/cm\textsuperscript{2}, achieving greater than 80\% diffraction efficiency. We also show that that beam pointing, divergence, and diffraction efficiency are stable while operating at 10 Hz.
These diffractive lenses are simple holograms, and the principles demonstrated here extended to other types of optics suggests that gaseous optics could enable arbitrary, damage-resistant manipulation of intense light for next-generation ultra-high-power lasers.

\end{abstract}

\maketitle

The world's highest energy and highest peak-power lasers, which produce up to several-megajoule nanosecond and ten-petawatt femtosecond pulses, respectively, have driven advances in inertial confinement fusion (ICF) \cite{AbuShawareb2022}, laboratory astrophysics \cite{fiuza_electron_2020,chen_perspectives_2023}, and the development of compact sources of intense radiation \cite{gibbon_harmonic_1996,dromey_coherent_2013,corde_femtosecond_2013,albert_applications_2016,Edwards2020} and accelerated particles \cite{tajima_laser_1979,Geddes2004,mangles_monoenergetic_2004,faure_laserplasma_2004,Esarey2009,Macchi2013,wang_quasi-monoenergetic_2013,miao_multi-gev_2022,gonsalves_petawatt_2019}. However, typical solid-state optics tolerate only 1-10 J/cm\textsuperscript{2} for nanosecond-duration pulses \cite{ristau_laser_2009,wang_laser-induced_2010} and 0.1-1 J/cm\textsuperscript{2} for sub-picosecond-duration pulses \cite{poole_femtosecond_2013}, so steering and focusing beams from these lasers requires meter-scale mirrors, gratings, and lenses---and optics cannot be placed near focus. 
Large (and both delicate and expensive) target-facing optics are also susceptible to non-optical damage, whether from neutrons and shrapnel in an ICF experiment, debris from solid-target interactions, or accelerated particle beams; this fragility restricts applications. 
Exploiting next-generation ultra-high-power lasers will require order-of-magnitude improvement in optical damage thresholds and optics that can survive unforgiving experimental environments.
The ability to focus light is fundamental for using lasers: without focusing, high energy and power cannot be turned into high fluence and intensity, so there is a specific need for damage-tolerant lenses. 

Gases and plasmas offer dramatically higher damage thresholds than solids and are therefore attractive media for ultra-robust optics \cite{andreev2006short,Milchberg2019,Malkin1999,leblanc_plasma_2017,wu_spatiotemporal_2025,hur_laser_2023,edwards_plasma_2022}. Gas optics are renewable---thus immune to cumulative damage---and have a higher threshold for avalanche ionization than solids \cite{bettis_correlation_1992}, increasing within-pulse damage tolerance. Plasmas are both renewable and already ionized, so, for femtosecond pulses, are limited primarily by the electron density distortions that occur at near-relativistic intensities \cite{edwards_plasma_2022}.
The refractive indices ($n$) of gases and plasmas deviate from unity ($n>1$ and $n<1$, respectively) by an amount related to their densities; shaping density therefore provides direct control over the phase of transmitted light, enabling transmission optics.
Refractive plasma and gas lenses---direct analogues of standard glass lenses---have been proposed \cite{ren_compressing_2001,hubbard_high_2002,palastro_plasma_2015} and created with plasma via the formation of a plasma channel by laser heating \cite{durfee_light_1993,katzir_plasma_2009,seemann_refractive_2023}, a capillary discharge \cite{gaul2000production,svirplys_attosecond_2025}, or the nonlinear refractive index \cite{kalmykov_dark-current-free_2011,kalmykov_dark-current-free_2012} and with gases using nozzles \cite{drescher_extreme-ultraviolet_2018} and temperature gradients or vortices to modify gas density in tubes \cite{berreman_b_1964,marcuse_analysis_1964,beck_thermal_1964,kaiser_measured_1968,notcutt_spinning_1988, notcutt_gas_1989,michaelis_gas-lens_1991}. Whether made in gas or plasma, refractive lenses are restricted to long focal lengths and are exquisitely sensitive to the density profile; minor inhomogeneities will substantially degrade the quality of focus. These lenses must also be relatively thick, resulting in nonlinear beam distortion at high powers, and are limited in diameter, the exact opposite of the thin (to minimize B-integral) and wide (to maximize total energy and power) aspect ratio most desirable for high-energy-laser optics. 
A more general solution for focusing high-power lasers therefore requires a different approach; we show here that diffraction, rather than refraction, provides a mechanism capable of delivering stable, robust optics with large diameters and millimeter-scale thicknesses.

Diffractive transmission gratings can be written into gas or plasma using the interference between two ``write" laser beams to impart a periodic refractive index modulation from which a third ``read" beam will diffract. Several mechanisms allow the creation of efficient gratings, including ionization, where gratings consist of alternating layers of gas and plasma \cite{suntsov_femtosecond_2009,Shi2011,Durand2012,edwards_control_2023,edwards_greater_2024}; ponderomotive forcing of fully ionized plasma \cite{lehmann_transient_2016,peng_nonlinear_2019}, which is closely related to cross-beam-energy-transfer \cite{Michel2009,Moody2012}; and gas heating, which can produce strong density modulations with nanosecond to microsecond lifetimes \cite{michine_ultra_2020,michel_photochemically_2024,ou_experimental_2025,ou_experimental_2025-1,oudin_piafs_2025,michine_large-amplitude_2024,matteo2025demonstration}.
This third method has been demonstrated in ozone-containing mixtures where ultraviolet light is strongly absorbed \cite{michine_ultra_2020}. When driven by two crossed ultraviolet write beams, ozone photodissociates into molecular and atomic oxygen in the constructive fringes of the write-beam interference pattern \cite{michel_photochemically_2024}, producing a periodic temperature variation that hydrodynamically evolves to form a transient density modulation, creating a diffraction grating \cite{michine_ultra_2020,michel_photochemically_2024,ou_experimental_2025,ou_experimental_2025-1,oudin_piafs_2025}.
Laser-driven gas gratings have much larger density fluctuations ($>$10\%) \cite{michine_ultra_2020,ou_experimental_2025-1} compared to transducer-driven acoustic gratings \cite{schrodel_acousto-optic_2024}, providing index modulations more than two-orders-of-magnitude larger ($<10^{-7}$ vs.\ $10^{-5}$-$10^{-4}$) and thus two-orders-of-magnitude thinner optics (5 mm vs.\ 49 cm) with higher efficiency, broader bandwidth, and weaker nonlinearity.
Furthermore, laser-driven gas gratings are simple holograms where the object beam is a plane wave \cite{eichler_laser-induced_1986,pochi_introduction_1993}.
It has been shown that the same mechanism can therefore be used to form more complex holographic optics \cite{edwards_holographic_2022,lehmann_plasma_2019,Edwards2024cleo}, one of which is a diffractive lens \cite{edwards_holographic_2022,lehmann_plasma_2019}. 

In this work, we demonstrate high-damage-threshold off-axis diffractive gas lenses that can steer and focus laser pulses with durations from femtoseconds to nanoseconds with greater than 80\% diffraction efficiency. 
Adjustment of the write-beam focal position allows tuning of the lens focal length.
We tested these lenses at fluences up to 35 J/cm\textsuperscript{2} for 532-nm, 5-ns pulses and intensities up to 5.5$\times$10\textsuperscript{13}~W/cm\textsuperscript{2} for 800-nm, 35-fs pulses.
These lenses produce near-diffraction-limited focal spots with good pointing stability at 10 Hz, making them suitable for high-energy and high-power lasers at high repetition rate.

\section{\label{sec:theory}Theory}

\begin{figure*}[t]
	\centering
	\includegraphics[width=\textwidth]{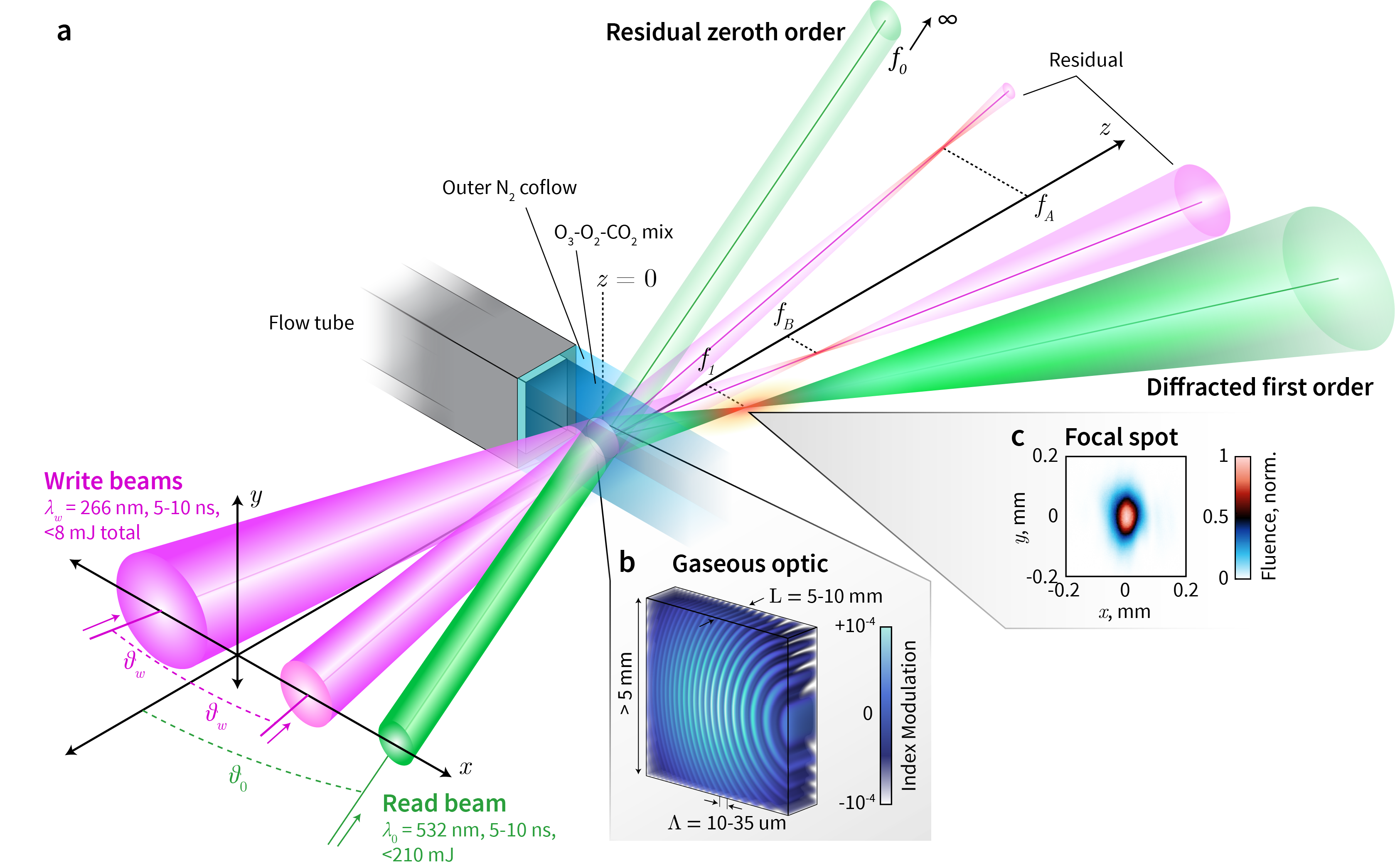}
	\caption{(a) Experimental configuration of an off-axis focusing diffractive gas lens. The ultraviolet write beams form a refractive index structure in an ozone-oxygen-carbon-dioxide gas flow. The resulting optic controls the focus of the diffracted read beam. (b) Simulated gas density profile, showing curved fringe pattern. (c) Experimental measurement of a diffracted beam focal spot.}\label{fig:Schematic}
\end{figure*}

A schematic of a diffractive gas lens is illustrated in Fig. \ref{fig:Schematic}. Two write beams of wavelength $\lambda_w$ propagate in the positive-$z$ direction with crossing half-angle $\theta_w$ in the $x$-$z$ plane. They intersect and fully overlap in a gas flow, where $z=0$ is set to be their intersection point and lies at the center of the flow. The two beams are focused at distances $f_A$ and $f_B$ from this intersection point along their respective propagation axes (approximately along $z$ when $\theta_w$ is small). 
When the resulting intensity interference pattern is imprinted as a refractive index structure in the gas, it forms a transmission holographic lens (Fig.~\ref{fig:Schematic}b). A read beam of wavelength $\lambda_0$, lying also in the $x$-$z$ plane, is most efficiently diffracted when incident at the Bragg angle, $\theta_\mathrm{B} = \arcsin{\left(\lambda_0/2n_0\Lambda\right)}$, corresponding to the local fringe period $\Lambda(x,y)$ and average refractive index $n_0$ \cite{pochi_introduction_1993}, resulting in diffraction to multiple orders with angles $\theta_m$ and focii $f_m$ given by,
\begin{equation} \label{eq:DiffractedAngle}
	\theta_m = \arcsin{\left(m\frac{\lambda_0}{n_0\Lambda}-\sin{\theta_0}\right)}
\end{equation}
\begin{equation} \label{eq:DiffractedFocus}
	f_m = \frac{f f_0}{f + mf_0},
\end{equation}
where integer $m$ is the diffraction order, $f_0$ is the incident read-beam focus, and $f = \left(\lambda_w/\lambda_0\right)\cdot f_A f_B/\left(f_A-f_B\right)$ is the focal length of the gas optic itself \cite{edwards_holographic_2022}.

If the crossing angle of the write beams is much larger than the difference in their divergence angles, the fringes are only slightly curved and the optic's diffraction efficiency---the fraction of incident energy that is diffracted in the first-order beam---is similar to that of a volume transmission grating. Just as for a grating, we can distinguish the Raman-Nath regime, where multiple diffraction orders are present, and the Bragg regime, where only first-order diffraction is significant, with the dimensionless parameter $\rho = \lambda_0^2/\Lambda^2 n_0 n_1$ \cite{moharam_criterion_1978}. The Raman-Nath and Bragg regimes correspond to $\rho<1$ and $\rho\gg 1$, respectively. Here, $n_1$ is the first-order Fourier component of the index modulation associated with the grating period $\Lambda$ \cite{pochi_introduction_1993}.
Under the conditions explored in this paper, $\rho$ was between 4 and 45, and diffraction into higher orders was generally small ($\ll1\%$), so we consider only the Bragg regime ($m =1$ in Eqs. \ref{eq:DiffractedAngle} and \ref{eq:DiffractedFocus}), where for light incident at the Bragg angle, the diffraction efficiency is:
\begin{equation} \label{eq:DiffractionEfficiency}
\eta = \sin^2\left( \frac{\pi n_1 L}{\lambda_0 \cos{\theta_\mathrm{B}}} \right),
\end{equation}
and the optic thickness ($L$) for efficient diffraction is:
\begin{equation} \label{eq:Thickness}
	\frac{2 n_1 L}{\lambda_0} \approx 1.
\end{equation}
As we will show here, being able to achieve $n_1 = 5\times 10^{-5}$ means that a 5-mm-thick gas optic can efficiently diffract 532-nm light.

\section{\label{sec:results}Results}

We constructed high-efficiency off-axis diffractive lenses in three different configurations that demonstrate (A) collimation of high-fluence nanosecond pulses, (B) tunable focusing, and (C) collimation of a broad-spectrum femtosecond pulse. The lenses were constructed by two ultraviolet write beams (266~nm, 5~ns, 2-4~mJ each) crossed in a premixed \ce{O3}-\ce{O2}-\ce{CO2} gas flow (1-5\% ozone, 20-60\% carbon dioxide).

\subsection{Experimental Realization of a Diffractive Focusing Optic}

\begin{figure*}
	\centering
	\includegraphics[width=\textwidth]{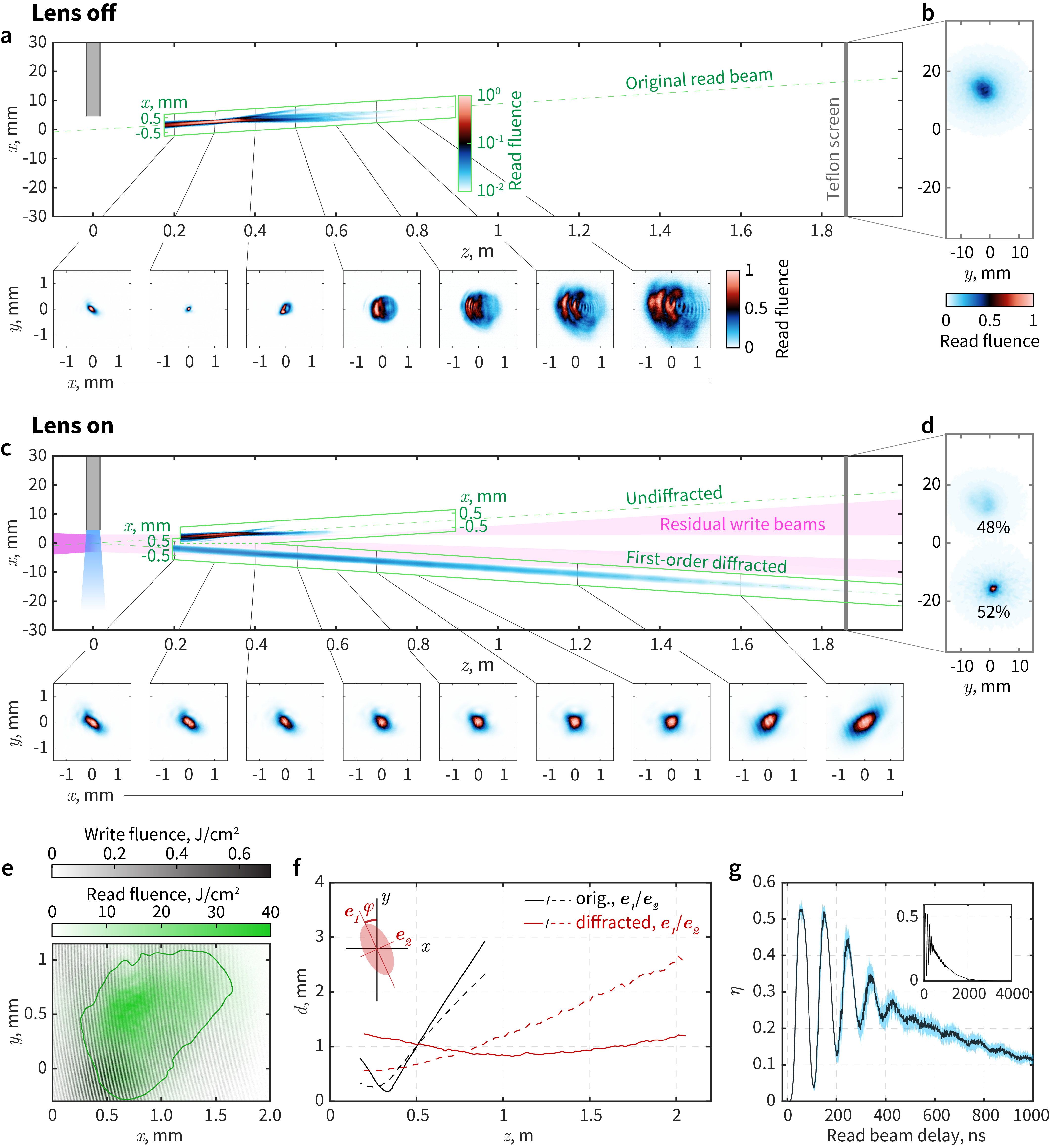}
	\caption{Experimental demonstration of a diffractive off-axis gas lens that collimates a focused 532-nm, 5~ns, 210~mJ read beam at a peak fluence above conventional solid-state damage thresholds. (a,c) Experimental read beam intensity slices in the x-z plane. The green axes outline the region in which the beams were imaged and indicate the scale of the transverse coordinate. (b,d) The original, undiffracted, and first-order diffracted beams imaged on a screen. (e) Fluence profiles of both the pump and probe beams in the gas. (f) Measured beam diameters as a function of distance after the optic in each beam's two principal axes. (g) Dependence of diffraction efficiency on relative timing between read and write beams for a focusing gas lens configuration in Fig. \ref{fig:AdjustableLens}(c). The shaded blue region is one standard deviation across 20 shots at each delay.}\label{fig:ExperimentalDemonstration}
\end{figure*}

Conventional solid-state lenses cannot collimate a high-energy laser beam at a small diameter because the high near-focus fluence will destroy any solid lens. Here, we made a gas lens capable of collimating a high-energy beam by crossing two write beams in the $x$-$z$ (horizontal) plane at a full angle $2\theta_w=$~0.6$^\circ$ with both beams having 3~mm beam diameters at their point of intersection in the gas flow (4.3\% \ce{O3}, 50\% \ce{CO2}). One write beam was collimated, and the other was focused 655~mm after the gas flow ($f_A =655$~mm, $f_B = \infty$). The high-energy read beam (532~nm, 210~mJ, 5~ns) was initially focused 280~mm after the gas lens ($f_0 = 280$~mm) with a 1.2~mm beam diameter at the lens, a 0.6$^\circ$ horizontal angle of incidence from the same side of the $y$-$z$ plane as the focused write beam, and a 1.5$^\circ$ vertical angle of incidence from the $x$-$z$ plane. With the gas optic off, the read beam propagated through a focal spot and expanded to 12~mm on a screen 1.86~m from the gas lens, as shown in Fig.~\ref{fig:ExperimentalDemonstration}(a,b).

When turned on, the gas optic diffracted 52\%$\pm$2\% of the incident read-beam energy into the first order, as shown in Fig.~\ref{fig:ExperimentalDemonstration}(c,d). The diffracted beam was steered by 1.0$^\circ$ in the $x$-$z$ plane and defocused about its optical axis. By fine-tuning the focal position of the focused write beam, we collimated the read beam with a constant transverse profile over two meters, as shown in Fig.~\ref{fig:ExperimentalDemonstration}(c). The diffracted beam was therefore more intense than the undiffracted beam on the Teflon screen shown in Fig.~\ref{fig:ExperimentalDemonstration}(d). The read and write beam profiles in the gas optic are shown in Fig.~\ref{fig:ExperimentalDemonstration}(e); the write beams formed curved interference fringes due to differing divergence. Note that the fluence of the read beam is almost two orders of magnitude higher than the fluence of the write beams; the energy required to form the optic is negligible compared to the energy that can be manipulated by it. 

With a peak read-beam fluence of 35~J/cm\textsuperscript{2} in the gas, this optical configuration surpasses the damage threshold of solid-state optics. Non-focusing diffraction gratings are expected to have laser-induced damage thresholds above 1~kJ/cm\textsuperscript{2} \cite{michine_ultra_2020}. Although the limited energy available in the read beam and the mm-scale diameters required to demonstrate focusing prevented higher fluence in this experiment, it is reasonable to expect that diffractive gas lenses can tolerate kJ/cm\textsuperscript{2} fluences. 

As shown in Fig.~\ref{fig:ExperimentalDemonstration}(f), direct imaging measurements of the diffracted beam on a camera chip indicate that it was collimated in its $\mathbf{e_1}$ principal axis over more than two meters. The two axes of the diffracted beam exhibited different focusing because the read beam was initially astigmatic and incident with an angle out of the $x$-$z$ plane \cite{young_zone_1972}, the write beams were spatially nonuniform, and holographic optics are generally prone to first-order (Seidal) aberrations  \cite{forshaw_imaging_1973}. The imperfections in the incident read and write beams are specific to our laser and could be resolved with a higher-performance system. For applications, the aberrations inherent to diffractive lenses can be compensated with an upstream deformable mirror \cite{hartley_wavefront_1997}.

The transient behaviour of non-focusing gratings in ozone gas has been studied in simulation \cite{michel_photochemically_2024} and measured experimentally \cite{michine_ultra_2020,michine_large-amplitude_2024,ou_experimental_2025}; we find here that an off-axis gas lens has similar temporal characteristics. The hydrodynamic response of the gas consists of two counter-propagating acoustic waves and a stationary entropy wave \cite{michel_photochemically_2024}. The superposition of these waves produces a density and refractive index structure, creating a transient optic (Fig.~\ref{fig:ExperimentalDemonstration}(g)). The hydrodynamic waves damp into bulk gas heating over several microseconds and the gas is fully replaced within a few milliseconds. Unlike non-focusing gratings, an off-axis zone plate has a period $\Lambda(x,y)$ that varies with $x$ and $y$, resulting in a mismatch of acoustic wave periods, although we find that this effect does not substantially change performance in an off-axis lens configuration where the variation of lens period is small compared to its average value. For the results shown in Fig.~\ref{fig:ExperimentalDemonstration}(c,d,f), the read beam was sent 52~ns after the write beams (at first maximum of $\eta$ in Fig.~\ref{fig:ExperimentalDemonstration}(g)).

We premixed \ce{CO2} into the gas flow because carbon dioxide creates a reaction pathway that quenches excited electronic states into translational states and has a proportionally greater change in refractive index than oxygen for the same density modulation \cite{michel_photochemically_2024}. These properties improved the energy efficiency of lens formation and the diffraction efficiency.

\subsection{Focal Length Tunability}

The focal length of a holographic lens can be tuned by modifying the focal positions of the write beams (Eq.~\ref{eq:DiffractedFocus}). To demonstrate control over the divergence of the diffracted read beam, we varied the focal position of one write beam and then measured the transverse profile of the diffracted beam as a function of longitudinal position (see Supplementary Information for beam configurations \cite{supplementary}). These measurements were conducted using a spatially cleaned, weakly diverging ($|\theta_d|<0.5$~mrad) read beam with a 0.5$^\circ$ incident angle from the same side of the $x=0$ plane as the adjustable write beam and a 2.7$^\circ$ angle downwards into the $x$-$z$ plane. Small adjustments to the beams' pointing, their relative delay, and the gas composition were made to maximize diffraction efficiency.

\begin{figure*}
	\centering
	\includegraphics[width=\textwidth]{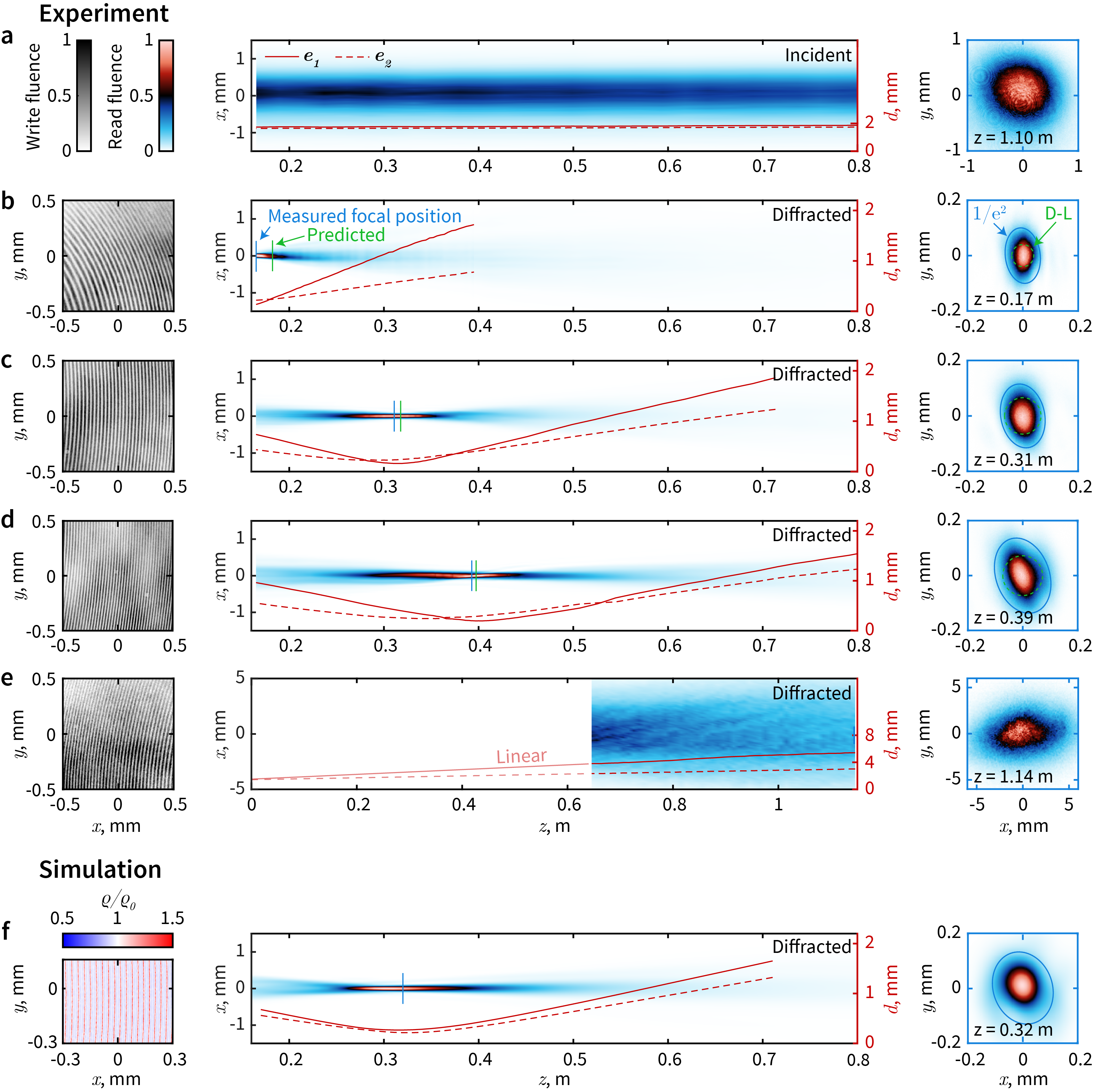}
	\caption{Control of the diffracted read-beam focusing by changing one of the write-beam focal positions. (a) Incident read beam in the $x$-$z$ plane (center) and $x$-$y$ plane far from the gas optic (right). (b-e) Measurements of the write-beam interference pattern (left) and the diffracted read-beam profile in the $x$-$z$ (center) and $x$-$y$ planes (right) in three different focusing configurations and a defocusing configuration, respectively. The predicted focal positions are calculated from Eq.~\ref{eq:DiffractedFocus} with the divergence of the incident beam taken into account. The transverse read-beam profiles (right) are taken at focus for (b,c,d). The solid blue line is the measured $1/e^2$-fluence contour and the dotted green line outlines a diffraction-limited focal spot for the same f-number. The read beam is imaged directly on a chip for (a-d) and on a screen for (e). (f) A simulation for the experimental configuration shown in (c) using paraxial linear propagation through a density structure predicted by the PIAFS hydrodynamic simulation. Diffraction efficiencies for (b-e) are 29\%, 50\%, 56\%, and 35\%, respectively.}\label{fig:AdjustableLens}
\end{figure*}

Figure~\ref{fig:AdjustableLens} shows measurements of the longitudinal and transverse profiles of the incident beam without the gas lens and the diffracted beam after the gas lens for three positive and one negative focal length. By changing the fringe pattern, we adjusted the focal spot from 16~cm to 40~cm after the gas optic (b-d), to 73~cm before the optic (e). Figure~\ref{fig:AdjustableLens}f shows calculated beam profiles from a paraxial propagation simulation of a lens profile calculated with the hydrodynamic code PIAFS (see Methods) in the same configuration as Fig.~\ref{fig:AdjustableLens}c; the hydrodynamic simulations overpredict the density modulation, but we find that using a density map at an earlier read-beam time delay correctly reproduces the focal position and elliptical beam shape of the diffracted beam.
The focusing properties of the optic originate from the fringe curvature: the fringe curvature decreases for increased lens focal length (b-d), and then flips sign in the defocusing case (e).
The diffracted focal spots are close to diffraction-limited and have little energy in their transverse wings. Equation~\ref{eq:DiffractedFocus} predicts the focal position to within about one Rayleigh length of the measured location.

\subsection{Diffraction of Femtosecond Pulses}

\begin{figure}[t]
	\centering
	\includegraphics[width=\columnwidth]{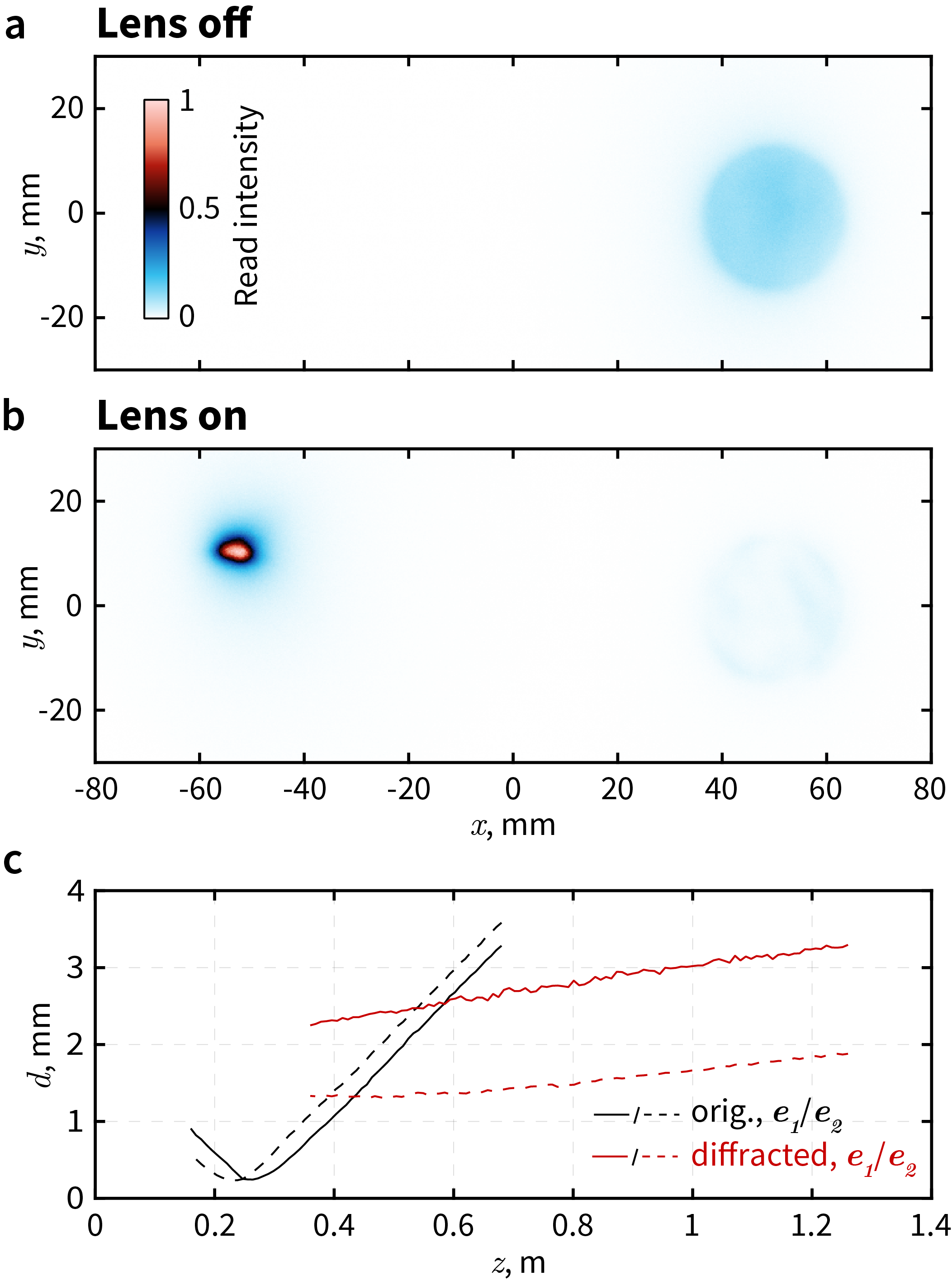}
	\caption{Diffraction of an initially focusing 800~nm, 35~fs, $\leq 80$~\si{\micro\joule} read beam in a collimating configuration through a gas lens. (a-b) The beams imaged on a screen with the lens on and off at $z=$~3.5~m. (c) Original and diffracted beam diameters as a function of distance after the gas lens.}\label{fig:Femtosecond}
\end{figure}

\begin{figure*}
	\centering
	\includegraphics[width=\textwidth]{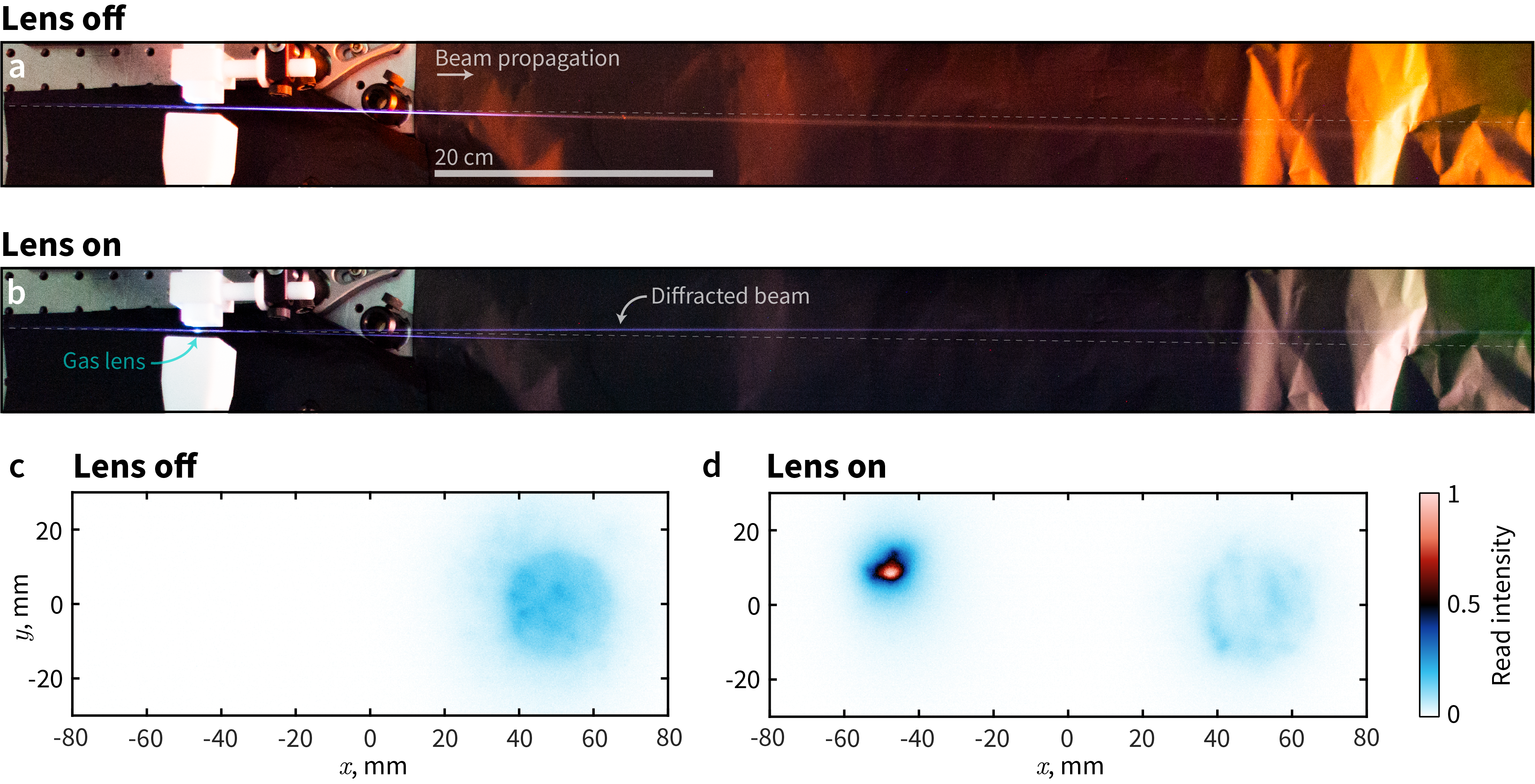}
	\caption{Collimation of an initially focusing 800~nm, 35~fs, 0.75~TW read beam. (a,b) Side-emission long-exposure (300 laser shots) photographs taken with a Digital Single-Lens Reflex (DSLR) camera from above the experimental setup. These images have been digitally enhanced in a spatially uniform manner to improve visibility, with identical processing applied to both. (c,d) Single-shot images of the beams on a Teflon screen at $z=3.5$~m with a $\lambda\geq700$~nm longpass filter.
    The read-beam energy was 30~mJ for the images captured by the DSLR and 28~mJ for measurements off the Teflon screen.}\label{fig:Femtosecond_HighEnergy}
\end{figure*}

To demonstrate the applicability of a gas lens to broad bandwidth, high peak-power femtosecond pulses, we changed the focusing of an 800~nm, 35~fs (45~nm FWHM bandwidth), 28~mJ read beam from a Ti:sapphire laser system. The incident read beam had an initial focal position 240~mm after the gas with a 2-mm diameter at the gas optic. The focal positions of the two write beams were configured to collimate the diffracted read beam. The larger Bragg angle of 800~nm light enabled the read and write beams to be coplanar.
We first characterized the performance of these lenses at low peak-powers ($\leq 3$~GW), where propagation through the lens and to the imaging screen would be linear. As shown in Fig.~\ref{fig:Femtosecond}(a,b), we measured a diffraction efficiency of 81\%$\pm$2\%, with the two axes of the diffracted beam having substantially lower divergence angles than those of the incident beam (Fig.~\ref{fig:Femtosecond}c).

In the same collimating configuration, we raised the peak power of the read beam to 0.75~TW, resulting in a peak intensity greater than $5.5\times10^{13}$W/cm\textsuperscript{2} in the lens.
%
%
We imaged the side-emission of the beams in air with and without the lens, and estimated the energy in all the beams by imaging the Teflon screen at $z=3.5$~m using calibrated filters.
As shown in Figs. \ref{fig:Femtosecond_HighEnergy}(a,c), the initial read beam ionized air at focus and contained 19~mJ of energy at the Teflon screen (68\% of incident energy).
With the lens turned on, Fig. \ref{fig:Femtosecond_HighEnergy}(d) shows that the diffracted read beam contained 13~mJ (46\% of incident energy) at a higher intensity at the Teflon screen than the original read beam. As shown in Fig. \ref{fig:Femtosecond_HighEnergy}(b), the side-emission photos with the lens turned on reveals a long multi-meter-long column of scattered light, consistent with the measured beam size at low power of a collimated diffracted beam.
The side-emission photographs were taken with room lights turned off, so differences in background light in Figs. \ref{fig:Femtosecond_HighEnergy}(a) and \ref{fig:Femtosecond_HighEnergy}(c) are largely due to differences in supercontinuum generated by the incident and diffracted beams.

In general, efficient diffraction of an ultrashort beam requires its spectral bandwidth to be within that of the optic. Non-focusing diffraction gratings operated at peak efficiency have a FWHM spectral bandwidth \cite{pochi_introduction_1993,edwards_plasma_2022}:
\begin{equation}
	\frac{\Delta \lambda}{\lambda_0} \approx 0.8 \frac{n_1}{\sin^2{\theta_B}}.
    \label{eqn:spectralbandwidth}
\end{equation}
This equation holds approximately true for an off-axis lens. 
In this configuration, with $\theta_B = 0.9^\circ$ and, assuming maximized efficiency at $L = 10$~mm to give $n_1 = 4\times10^{-5}$, Eq.~\ref{eqn:spectralbandwidth} predicts $\Delta \lambda = 104$~nm, in agreement with the observation of efficient diffraction across the full bandwidth of our femtosecond pulse.

\subsection{Stability}
\begin{figure*}
	\centering
	\includegraphics[width=\textwidth]{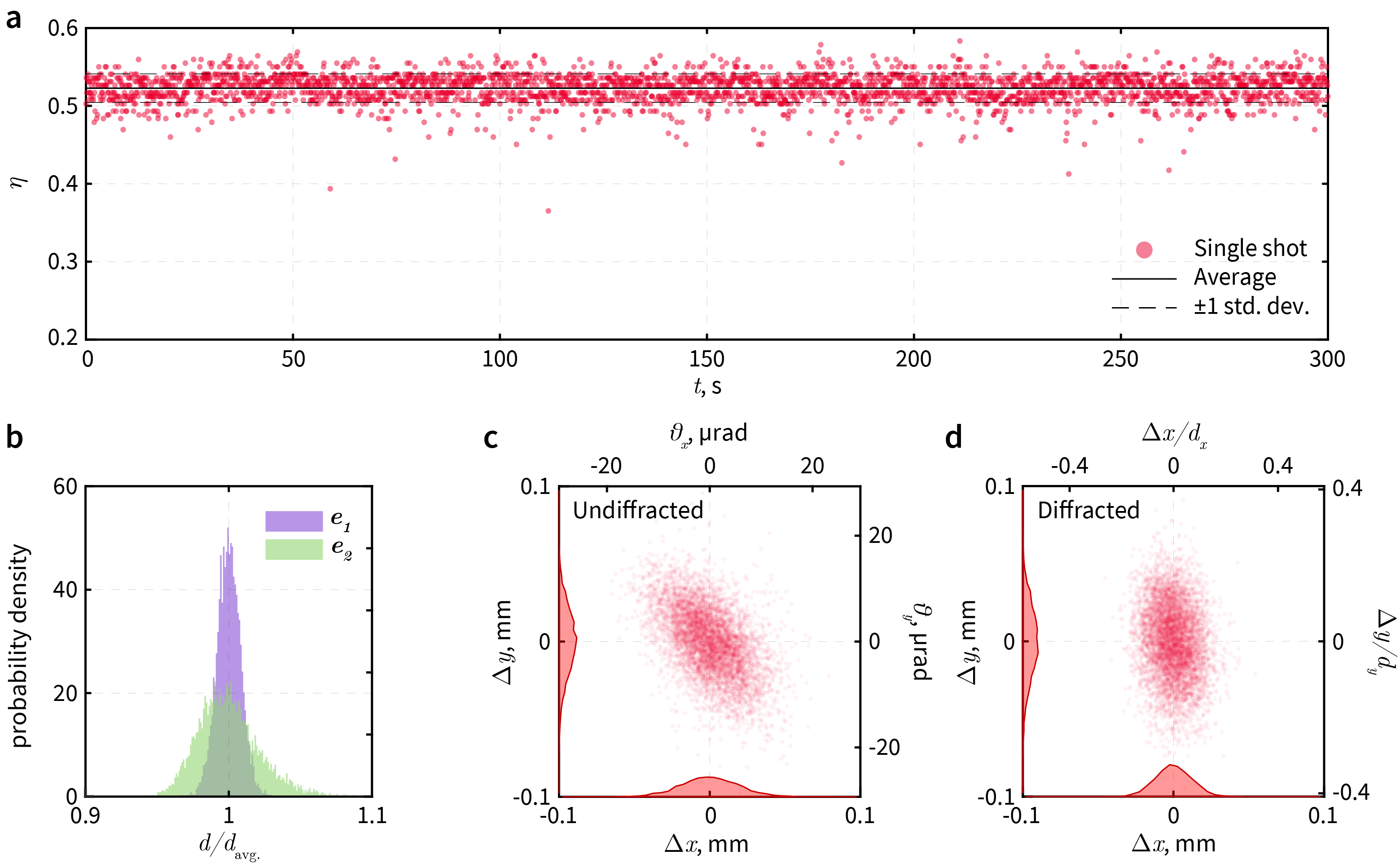}
	\caption{Stability of a gaseous lens. (a) Diffraction efficiency for five minutes of operation at the same configuration in Fig. \ref{fig:ExperimentalDemonstration} at a 10~Hz repetition rate. (b) Distribution of relative beam diameters at the focal plane with a focusing gas lens (same configuration as Fig. \ref{fig:AdjustableLens}c). The focal spot was \SI{173}{}$\pm$\SI{1}{\micro\meter} and \SI{245}{}$\pm$\SI{5}{\micro\meter} along the $\mathbf{e_1}$ and $\mathbf{e_2}$ axes which were approximately aligned with the $\mathbf{x}$ and $\mathbf{y}$ coordinate axes, respectively. (c,d) Distribution of beam centers of the incident and diffracted read beam, respectively, at the focal plane of the diffracted beam. Data in (b-d) represents 6000 shots (10 minutes at 10~Hz).}\label{fig:Stability}
\end{figure*}

To check the stability and reproducibility of the gas lenses, we measured diffraction efficiency for five minutes at 10 Hz (3000 shots) in the high-fluence collimating configuration shown in Fig.~\ref{fig:ExperimentalDemonstration}. Here, diffraction efficiency was calculated as the ratio of single-shot diffracted energy to the average incident energy. Figure~\ref{fig:Stability}(a) shows that the diffraction efficiency was 52$\pm$1.8\% (3.5\% relative standard deviation) with random shot-to-shot fluctuations but no long-term trend. The total energy in both write beams was 7.8$\pm$0.8~mJ (10\%) and the incident read-beam energy was 210.9$\pm$4.4~mJ (2.1\%). 
Diffraction efficiency was observed to be more stable than the write-beam energy. This can be explained by noting that when Eq.~\ref{eq:Thickness} is satisfied, we are near a local maximum of Eq.~\ref{eq:DiffractionEfficiency}, and small changes in diffraction efficiency ($\delta\eta$) depend on the square of small changes in the refractive index modulation ($\delta n_1$), i.e.~$\delta\eta \sim -\left(\pi L / \lambda \cos\theta_B\right)^2\left(\delta n_1\right)^2$. A gas lens operating near peak efficiency can therefore be relatively robust to small write-laser or gas-flow fluctuations.

We characterized the beam size and pointing stability by imaging the incident and diffracted read beams in the focusing configuration shown in Fig.~\ref{fig:AdjustableLens}(c) for ten minutes. As shown in Fig.~\ref{fig:Stability}(b), the diffracted beam waist is stable to within 0.8\% and 2.1\% in its two principal axes. Figures~\ref{fig:Stability}(c) and \ref{fig:Stability}(d) show the single-shot variation in transverse beam position of the incident and diffracted beams, both measured at the $z$-position of the diffracted focal spot. The selective stabilization of horizontal beam pointing is associated with the smaller angular bandwidth of the lens in the horizontal direction.

\section{Discussion}
Diffractive gas lenses may be suitable for a variety of high-energy and high-peak-power laser facilities and experiments. Consider, for example, the final focusing lenses of the National Ignition Facility (NIF) beamlines, that each handle a 40 cm $\times$ 40 cm, 13 kJ pulse. At an expected damage threshold around 1.5 kJ/cm\textsuperscript{2} \cite{michine_ultra_2020,ou_experimental_2025}, a gas lens would need only be 3 cm $\times$ 3 cm, requiring 1-2 J of ultraviolet writing energy. In addition to the high damage threshold and neutron, x-ray, and debris resistance of these lenses, their finite lifetime may, if appropriately tuned, also provide backscatter protection. Gas lenses of this type are therefore a potential answer to the unsolved final-optics challenge of high-repetition-rate inertial fusion energy facilities, where beam energies and shot repetition rates are expected to be much higher than those at NIF \cite{chapman_investigation_2019,schultz_ife_2001,suratwala_optic_2025}.

A low-density off-axis lens that reforms for each laser shot is also be beneficial for high-peak-power laser experiments where beam reshaping close to target is desirable. For instance, laser-plasma electron accelerators have demonstrated GeV electron energies \cite{miao_multi-gev_2022,gonsalves_petawatt_2019}, but producing TeV electron beams would require staging many individual acceleration sections together to counteract depletion of the driving laser pulse and dephasing between the laser and the electron beam \cite{luo_multistage_2018,steinke_multistage_2016,schroeder_physics_2010,albert_2020_2021}. Staging requires changing the direction of a high-power driving beam near focus in order to align it with the electron beam using an optic that the electron beam can travel through; a gas lens is an alternative to the plasma mirrors currently in use. 
More broadly, gas lenses allow table-top experiments with high-power beams at large f-number. A gas lens can manipulate high-energy beams much closer to focus than a glass lens, so reaching large spot sizes and long Rayleigh ranges does not require long and straight laboratory halls. The extension of gas optics to holographic lenses also suggests that more complex structuring of high-power light will be feasible. 

Here, we have experimentally demonstrated efficient diffractive gas lenses via UV-induced photodissociation of ozone. We have shown that these optics can collimate initially focusing nanosecond to femtosecond read beams at fluences and intensities well above solid-state damage thresholds and that the lens focal length can be tuned by changing the relative focal positions of the two write beams. The process of forming the optic is energy efficient; 100~mJ/cm\textsuperscript{2} write beams can manipulate read beams above 1~kJ/cm\textsuperscript{2}. We measured stable beam pointing and diffraction efficiencies that fluctuate less than the write-beam energy. These characteristics show the potential of using diffractive gas lenses for the control of high-energy lasers.

\begin{acknowledgments}
This work was supported by National Science Foundation Grant No.~PHY-2308641, NNSA Grant No.~DE-NA0004130, and the Laboratory Research and Development Program at LLNL under Project Tracking
Code No. 24-ERD-001. Lawrence Livermore National Laboratory is operated by Lawrence Livermore National Security, LLC, for the U.S. Department of Energy, National Nuclear Security Administration under Contract No. DE-AC52-07NA27344.
\end{acknowledgments}

\section{Appendix: Methods}

\subsection{Experimental Setup}

\begin{figure*}[ht]
	\centering
	\includegraphics[width=0.98\textwidth]{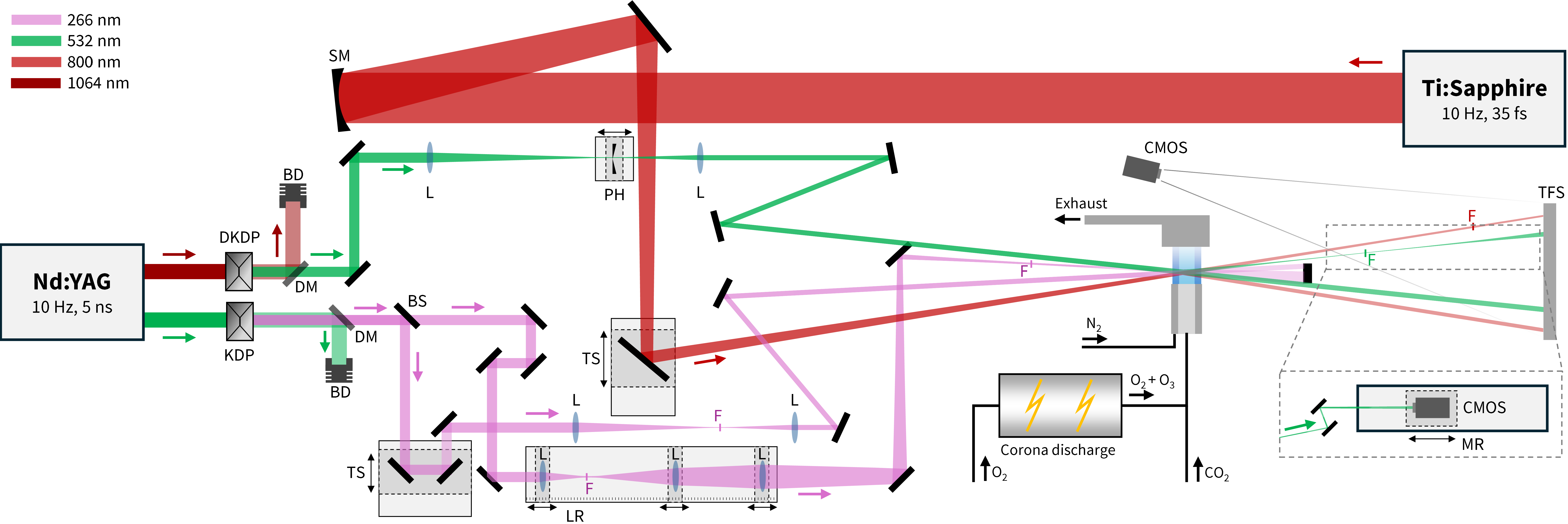}
	\caption{Experimental beamline. Abbreviations are potassium dihydrogen phosphate crystal (KDP), potassium dideuterium phosphate crystal (DKDP), dichroic mirror (DM), beam dump (BD), beam splitter (BS), lens (L), pinhole (PH), translation stage (TS), linear rail (LR), motorized rail (MR), Teflon screen (TFS), and complementary metal-oxide-semiconductor camera (CMOS).}\label{fig:beamline}
\end{figure*}

Figure \ref{fig:beamline} depicts the experimental setup used throughout this work. We used two electronically synchronized Nd:YAG lasers, one frequency quadrupled to 266~nm and the other frequency doubled to 532~nm. The 266~nm beam was split to provide the two write beams (5~ns, $\leq$\SI{7.8}{\milli\joule} total) and the 532~nm beam was used as the nanosecond read beam (532~nm, 5~ns, $\leq$\SI{220}{\milli\joule}). A Ti:Sapphire CPA-based laser system provided the femtosecond read beam (800~nm, 35~fs, $\leq$30~mJ). The write beams were incident on the gas flow with a full crossing angle ranging from 0.4$^\circ$ to 1.0$^\circ$. Both beams had approximately a 3 ~mm beam diameter in the gas. One of the write beams was collimated, and the other had a variable divergence half-angle ranging from \SI{-4.2}{\milli\radian} (f-number: -120) to \SI{4.5}{\milli\radian} (f-number: 110). The long-pulse read beam was used in two configurations: a low-energy, spatially cleaned, collimated beam with a 2~mm beam diameter and a full-energy, focused beam with an f-number of 230 and a 1.2~mm beam diameter in the gas. The femtosecond read pulse was focused with a spherical mirror, giving it an f-number of 110 and a 2~mm beam diameter at the gas. We controlled the relative timing between the read and write pulses with a digital delay generator. The final beam pointing and timing were fine-tuned during experiments to maximize diffraction efficiency.

Experiments were conducted at atmospheric temperature and pressure. The gas flow tube consisted of two nested rectangular channels--the inner channel contained a mixture of ozone ($<$10\%), oxygen, and carbon dioxide ($<$50\%) with a flow velocity near 1~m/s. The outer channel contained a co-propagating flow of nitrogen to mitigate turbulent shear mixing at the ozone interface (the write beams were not sensitive to the quality of the nitrogen-air interface). The thickness of the gas optics was controlled by the flow tube width; thickness of 5 mm and 10 mm were both used. The height of the flow ($y$) was 10 mm. Ozone was generated from pure oxygen in a discharge generator, then mixed with carbon-dioxide to reach the desired concentrations.

For most configurations, the diffraction efficiency was measured by directly imaging all diffracted orders on the same CMOS camera sensor. The diffraction efficiency was calculated as the sum of counts (after background subtraction) in the first-order diffracted beam divided by the total counts in all orders, which was confirmed to be equivalent to the total counts of the incident read beam ($>$99.9\% transmission through the gas flow). For experiments with the femtosecond read pulse (Fig.~\ref{fig:Femtosecond}) and in the defocusing configuration of the nanosecond read pulse (Fig.~\ref{fig:AdjustableLens}(e)), efficiency was measured by measuring scattering from a Teflon screen due to the larger diffraction angles and thus beam sizes. 

Diffraction efficiency with the high-fluence nanosecond read beam was calculated as the ratio of diffracted to incident laser energy as measured with an energy meter. The beam profiles shown in Fig.~\ref{fig:ExperimentalDemonstration} were collected at low energy; as the energy was increased, the far-field read-beam profile was observed to remain constant.
For experiments with the high-power femtosecond read beam, the energy in each beam was estimated from imaging a Teflon screen with calibrated filters. A 700-nm longpass filter was also used on the imaging camera. The transmission values for all filters were calibrated on the femtosecond read beam except one which was calibrated using an 830~nm continuous-wave diode laser. Peak read-beam fluences were estimated by imaging the beam profile at low energy.

All beam size measurements follow the ISO 11146-1:2021 standard; for a Gaussian beam, this definition is equivalent to the $1/e^2$ intensity diameter. Beams smaller than $\sim$2~mm were directly imaged on a CMOS camera chip (4.97$\times$3.73~mm sensor size, \SI{3.45}{\micro\meter} pixel size), while larger beams were imaged on a Teflon screen. The position of the imaging setup was scanned along each beam's propagation direction; at each position, 10 to 50 single shot images of the beam and 10 background images were captured. To reduce salt and grain noise, each image was processed with a square median filter (3~px size for beams directly on a chip, 5~px filter for beams on a Teflon screen). Beam properties were determined on each single-shot image independently; only the final measurements were averaged (ex. average beam diameter). Single-shot images of the beam at different $z$-positions were recentered, averaged, and then stacked to generate a 3D fluence model of the beam; slices in the $x$-$z$ plane, as shown in Fig.~\ref{fig:AdjustableLens} for example, were produced by plotting the $y=0$ plane of this volume.

Beam incidence angles were estimated by measuring their transverse position on a ruled beam block 76~cm before the gas flow. To image the write-beam interference pattern and the position of the read beam relative on the optic, all the beams were sampled directly upstream of the gas optic and imaged on a camera chip at the same distance along the propagation direction as the gas flow. The ozone concentration was measured via linear attenuation of a continuous-wave ultraviolet ($\lambda=$~266~nm) beam passing through the gas flow and the known cross-section of ozone at 266 nm. From the Beer-Lambert law, the ozone number density is,
\begin{equation}
	N_\mathrm{O_3} = \frac{1}{\sigma L}\ln{\left(\frac{I_0}{I_1}\right)},
\end{equation}
where $\sigma = 9.35\cdot10^{-22}$~m\textsuperscript{2} is the ozone absorption cross-section at 266 nm \cite{gorshelev_high_2014}, and $I_0$ and $I_1$ are the incident and transmitted light intensities, respectively.

The side-emission from the high-power read beam in air was captured by a Digital Single-Lens Reflex (DSLR) camera (Nikon D5200 with a Nikon AF-S DX NIKKOR 18-140 mm f/3.5-5.6G ED VR lens). Photographs were captured as raw image files and later processed in Adobe Lightroom with identical settings applied to all pairs of images with the lens on and off.

\subsection{PIAFS Simulation}

Numerical simulations in Fig. \ref{fig:AdjustableLens}(f) were performed with PIAFS \cite{oudin_piafs_2025}, a hydrodynamic simulation code that couples nonlinear fluid dynamics with chemistry. It relies on a conservative, high-order finite-difference scheme to solve the two-dimensional Navier–Stokes equations on a Cartesian grid, while simultaneously integrating the chemical reaction equations that describe the evolution of local species concentrations.

Experimental data were used to define the initial conditions. We used the measured write-beam fluence map to determine the UV-laser fluence and thus deposited energy in the gas at each position in $x$, $y$, and $z$. PIAFS then simulates the subsequent evolution of the gas for a series of two-dimensional slices along the depth of the grating. The density modulation was then converted into its corresponding refractive index and injected into a paraxial propagation code to simulate the propagation of the probe.

The simulation parameters match those of the experiments: [\ce{CO2}]=0.5, [\ce{O3}]=0.041, a grating length of 5 mm along the z-axis, a pulse duration of 5~ns, and an average fluence of 125 mJ/cm\textsuperscript{2}.
To isolate the focusing effect of the curved interference fringes from variations in diffraction efficiency due to differences between the simulated and experimental refractive index modulation amplitudes, we chose the time delay between the read and write beams such that $2n_1L/\lambda_0<1$. In these simulations, this condition was satisfied with a delay of 25~ns.
The computational grid consisted of 2048 cells along $x$ and 1024 along $y$. Since the fringes fall mostly in the $y$-$z$ plane, the strongest hydrodynamic response was in $x$, justifying the higher resolution in the $x$ direction. This setup provided 23 cells per grating period.

\subsection{Paraxial Simulations}

The paraxial wave equation can describe monochromatic light propagation at small transverse angles through a refractive index that varies slowly in the propagation direction:
\begin{equation} \label{eq:paraxial}
	\left(\nabla_\perp^2 + 2ik_0\partial_z\right)E(\Vec{r}) = 2k_0^2E(\Vec{r})\frac{n(\Vec{r})-n_0}{n_0},
\end{equation}
where $k_0$ is the wavenumber and $n_0$ is the background refractive index \cite{michel_introduction_2023}. We simulated the propagation of read beams through the gas optic using a refractive index map calculated with PIAFS, although similar behavior could be observed by modeling the gas optic as having a refractive index modulation proportional to the write-beam intensity:
\begin{equation}
	n(\Vec{r}) - n_0 = \delta n \frac{|E(\Vec{r})|^2}{\max{|E(\Vec{r})|^2}}
\end{equation}
with the index modulation, $\delta n$, fit to approximately match the experimental diffraction efficiency. The simulation also assumes that the incident read beam was circular, Gaussian, and aberration-free. Differences between simulated and experimental diffracted beam profiles are likely due to deviations from these assumptions.

\bibliography{refs_export}

\clearpage

\pagebreak
\widetext
\setcounter{equation}{0}
\setcounter{figure}{0}
\setcounter{table}{0}
\setcounter{page}{1}
\setcounter{section}{0}
\makeatletter
\renewcommand{\theequation}{S\arabic{equation}}
\renewcommand{\thefigure}{S\arabic{figure}}
\renewcommand{\bibnumfmt}[1]{[S#1]}
\renewcommand{\citenumfont}[1]{S#1}
\renewcommand{\thesubsection}{S\arabic{section}.\arabic{subsection}}


\title{Supplementary Material: Focusing Terawatt-Scale Lasers Using Holographic Gaseous Lenses}
\maketitle


To further validate our modeling of an off-axis holographic lens, we measure the spatial and temporal behaviour of higher-order diffracted modes from a gas lens. We also provide supplementary photographs of the side-emission of the high-power read beam in air after collimation by the gas lens, and the experimental parameters required to reproduce our results.

\section{Higher-order diffraction}

\begin{figure}[!ht]
	\centering
	\includegraphics[width=0.49\textwidth]{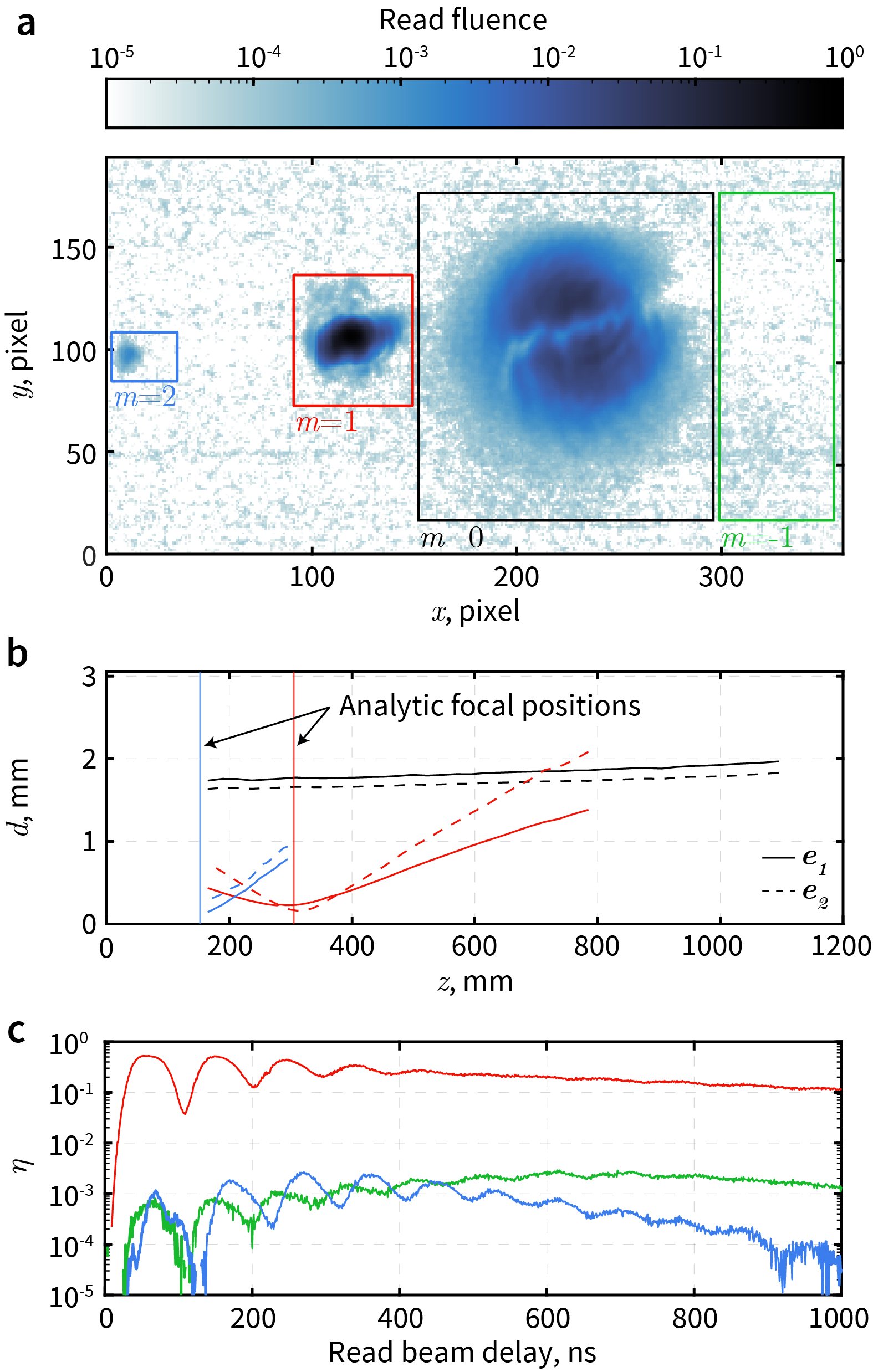}
	\caption{Experimental measurements of higher-order diffraction modes from a focusing gas lens configuration for $\rho\approx20$. (a) Three orders of diffracted beams and the undiffracted beam imaged directly on a camera after the gas lens. (b) Measured beam diameters as a function of position. Analytic focal positions are from Equation 2 from the main text. (c) Diffraction efficiency as a function of read beam delay. }\label{fig:HigherOrderDiffraction}
\end{figure}

The range of feasible density modulations (and $n_1$) at atmospheric conditions and modulation periods given the beam incidence and divergence angles puts the hologram into the Bragg regime. Consistent with this theory, we observed higher-order diffracted beams to contain much less energy ($\ll$~1\%) of the first-order diffracted beam as shown in Figure \ref{fig:HigherOrderDiffraction}(a). To characterize the behaviour of these modes, we measured their beam sizes as a function of position after the lens and their diffraction efficiency as a function of read beam delay. As shown in Figure \ref{fig:HigherOrderDiffraction}(b), the $m=2$ mode was focused tighter than the $m=1$ mode, and both of their measured focal positions are follow predictions from Equation 2 in the main text. The $m=-1$ mode was defocused through the gas lens, so only its total energy (sum of pixel values) could be measured rather than a beam profile. As shown in Figure \ref{fig:HigherOrderDiffraction}(c), the diffraction efficiencies of all orders oscillate in time and the higher orders are 2-3 orders of magnitude less energetic than the first-order diffracted beam. The higher order modes diffract off of different spatial periods of the refractive index structure, so as acoustic waves of non-constant $\Lambda(x,y)$ cross the region where the read beam is incident, the higher order modes oscillate with a slightly different temporal period. The different orders also maximize their efficiency at a later time than the first-order diffracted beam.

\clearpage

\section{Photographs of Diffraction at High-Power}

\begin{figure*}
	\centering
	\includegraphics[width=\textwidth]{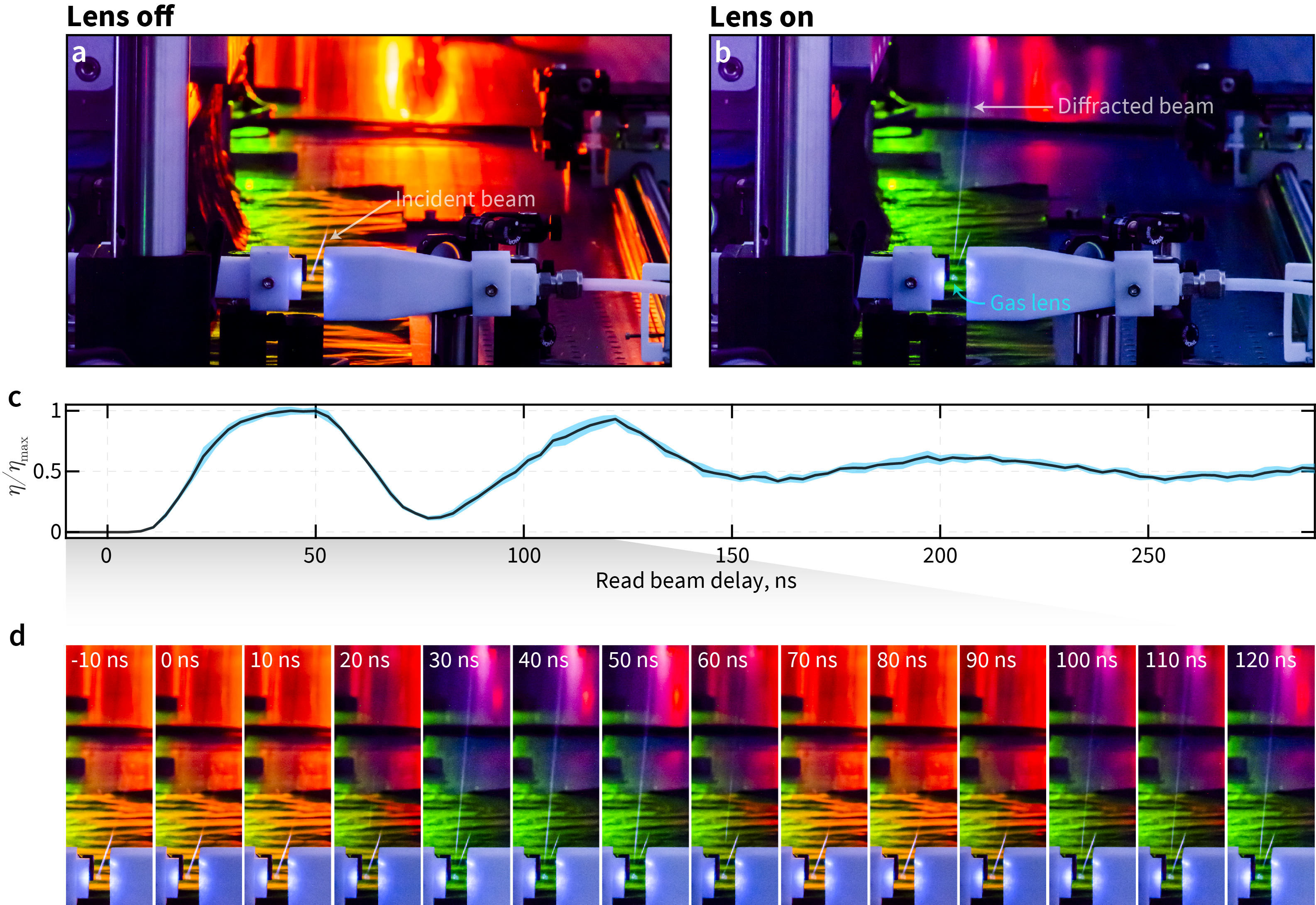}
	\caption{Photographs of collimation with a gas lens of a terawatt-class 800-nm femtosecond read beam propagating in air. (a,b,d) Long-exposure (300 laser shots) DSLR photographs taken with the lens turned (a) off at $t=-10$~ns, (b) on at $t=40$~ns, and (d) at varying time delay. (c) Energy in the diffracted beam as a function of read beam time delay. The read-beam peak power was 0.67~TW for the images in (a,b,d) and 0.75~TW for the data in (c). The green on the left is leaked light from an amplifier pump laser in the Ti:Sapphire laser system.}\label{fig:FemtosecondDelayScan}
\end{figure*}

In addition to the side-emission photographs taken from above the experimental setup in Fig. \ref{fig:Femtosecond_HighEnergy}, we also captured long-exposure photographs from a shallower angle to better visualize the angular separation of the incident and diffracted read beams. As shown in Fig. \ref{fig:FemtosecondDelayScan}(b), the gas lens collimates the read beam and creates a relatively long column of scattered light compared to the strong ionization at focus of the original read beam in Fig. \ref{fig:FemtosecondDelayScan}(a). Figures \ref{fig:FemtosecondDelayScan}(c) and \ref{fig:FemtosecondDelayScan}(d) show that the strength of the scattered light in the path of the diffracted beam captured by the DSLR camera is strongest at the time delays corresponding to peak efficiency of the gas lens.

\onecolumngrid

\clearpage

\onecolumngrid

\section{Experimental Parameters}

\begin{table}[ht]
\caption{Physical and Computational Parameters}
\label{tbl:params}
\begin{ruledtabular}
\begin{tabular}{l c c c  c c c  c c c  c c c  c c c}
\noalign{\smallskip}
{\bf Figure} & \multicolumn{4}{l}{\bf Write Beams\footnote{All configurations use a pair of write beams with $\lambda_w = 266$ nm, $\tau = 5$ ns, and $d = 3\pm1$ mm at their intersection in the gas flow.}} & \multicolumn{8}{l}{\bf Read Beam\footnote{The pulse duration of the 532-nm read beam was 5-10 ns. The pulse duration of the 800-nm read beam was 35 fs.}} & \multicolumn{3}{l}{\bf Gas Flow}\\
\noalign{\smallskip}
 &
$E$\footnote{Combined energy of both imprint beams.} &
$2\theta_w$\footnote{All angles are $\pm0.1^\circ$.} &
$f_A$\footnote{Focal positions are relative to the position of the gas lens and were measured using sparking on an alignment card to within 1 cm. Positive focal positions are after the gas lens.} &
$f_B$ &
$\lambda_0$ &
$E$ &
$\theta_0$ &
$\theta_y$ &
$d$ &
$f_0$ &
$\Delta \tau$\footnote{Write-read delay is based on pulse maxima. Positive delay indicates read beam arrival after the write beam.} &
S.C.\footnote{Spatial cleaning. Indicates whether read beam was spatially cleaned with a pinhole. See Fig. \ref{fig:beamline}.}&
$L$ &
[O$_3$]\footnote{Uncertainty in specific ozone concentration values is $\pm0.1\%$. Where a range is indicated, precision measurements were not collected.} &
[CO$_2$]\footnote{Uncertainty in specific carbon dioxide concentration values is $\pm10\%$. Where a range is indicated, precision measurements were not collected.} \\
\noalign{\smallskip}
 & (mJ) & ($^\circ$) & (cm) & (cm) & (nm) & (mJ) &($^\circ$) & ($^\circ$)& (mm) & (cm) & (ns) & & (mm) & & \\
\noalign{\smallskip}
\hline
\noalign{\medskip}
1c, 3b & $6.3$ & $0.4$ & $-36$ & $\infty$ & $532$ & $<0.1$ & -0.7 & $2.7$ & $2.0$ & $\infty$ & $60$\footnote{Approximate value} & Y & 5 & 1-5\% & 50\% \\ 
\noalign{\smallskip}
2a-f & $6.4$ & $0.6$ & $66$ & $\infty$ & $532$ & $210$ & -0.6 & $1.5$ & $1.2$ & $28$ & $88$ & N & 10 & 1-5\% & 50\% \\ 
\noalign{\smallskip}
6a & $7.8$ & $0.6$ & $66$ & $\infty$ & $532$ & $210$ & -0.6 & $1.5$ & $1.2$ & $28$ & $<100$ & N & 10 & 1-5\% & 30\% \\ 
\noalign{\smallskip}
2g, 3a, & $5.5$ & $0.9$ & $-61$ & $\infty$ & $532$ & $<0.1$ & -0.5 & $2.7$ & $2.0$ & $\infty$ & 57 & Y & 5 & 4.3\% & 50\% \\ 
3c, 6b-d,\\S1\\
\noalign{\smallskip}
3d & $4.6$ & $0.6$ & $-71$ & $\infty$ & $532$ & $<0.1$ & -0.3 & $2.7$ & $2.0$ & $\infty$ & 57 & Y & 5 & 1-5\% & 50\% \\ 
\noalign{\smallskip}
3e & $3.9$ & $0.4$ & $42$ & $\infty$ & $532$ & $<0.1$ & -0.7 & $2.7$ & $2.0$ & $\infty$ & 57 & Y & 5 & 1-5\% & 50\% \\ 
\noalign{\smallskip}
4 & $\leq10$ & $0.5$ & $70$ & $\infty$ & $800$ & $<0.1$ & -0.9 & $0.0$ & $2.0$ & $24$ & $<100$ & N & 10 & 1-5\% & 50-80\%\\ 
\noalign{\smallskip}
5, S2 & $\leq10$ & $0.5$ & $70$ & $\infty$ & $800$ & $25-30$\footnote{Exact values specified in main text.} & -0.9 & $0.0$ & $2.0$ & $24$ & $<100$ & N & 10 & 1-5\% & 50-80\%\\ 
\end{tabular}
\end{ruledtabular}
\end{table}

\twocolumngrid

\end{document}